\documentclass[preprint,12pt]{elsarticle}

\usepackage{amssymb}
\usepackage{graphicx}

\journal{Computational Materials Science}

\begin{document}

\author[INT]{M. J. Winiarski\corref{cor1}}
\address[INT]{Institute of Low Temperature and Structure Research, Polish Academy of Sciences, Ok\'olna 2, 50-422 Wroc\l aw, Poland, EU}

\title{The band-gap of Tl-doped gallium nitride alloys}

\begin{abstract}
Structural and electronic properties of hypothetical zinc blende Tl$_x$Ga$_{1-x}$N alloys have been investigated from first principles. The structural relaxation, preformed within the LDA approach, leads to a linear dependence of the lattice parameter {\it a} on the Tl content $x$. In turn, band structures obtained by MBJLDA calculations are significantly different from the corresponding LDA results. The decrease of the band-gap in Tl-doped GaN materials (for $x<0.25$) is predicted to be a linear function of $x$, i.e. 0.08~eV per atomic \% of thallium. The semimetallic character is expected for materials with $x>0.5$. The obtained spin-orbit coupling driven splitting between the heavy-hole and split-off band at the $\Gamma$ point of the Brillouin zone in Tl$_x$Ga$_{1-x}$N systems is significantly weaker when compared to that of Tl-doped InN materials. 
\end{abstract}


\maketitle

\section{Introduction}

Semiconducting group-III nitrides are well known materials used for fabrication of optoelectronic devices \cite{Nature}. Parent compounds adopting the wurtzite phase, with direct band-gaps starting from 0.65-0.69 eV for InN \cite{PhysStatSolB_229_R1,APL_83_4963,JApplPhys_94_4457}, through 3.50-3.51 eV for GaN \cite{JApplPhys_83_1429, JApplPhys_94_3675}, up to 6.1 eV for AlN \cite{JApplPhys_94_3675}, allow a design of ternary alloys within a wide interval of available wave lengths. A further band-gap engineering, leading to the infrared devices, could be realized by a doping with elements exhibiting a high atomic number.

Although some Tl-doped III-V group semiconductors have already been obtained, e.g. Tl$_x$In$_{1-x}$P \cite{JApplPhys_81_1704}, Tl$_x$Ga$_{1-x}$As, and Tl$_x$In$_{1-x}$As alloys \cite{PhysRevB_72_125209, JCrystGrow_237_1495}), the synthesis of such nitrides has not been reported yet. Meantime, the structural and electronic properties of hypothetical Tl$_x$Al$_{1-x}$N \cite{TlAlN1,TlAlN2}, Tl$_x$Ga$_{1-x}$N \cite{TlGaN}, and Tl$_x$In$_{1-x}$N \cite{TlInN} systems were investigated only from first priciples.

Standard DFT calculations within the local density (LDA) and generalized gradient (GGA) approximations for TlN predicted a semimetallic character of this compound with the zero band-gap \cite{TlN1,TlN2,TlN3,TlN4}. The recent fully relativistic study within the modified Becke-Johnson (MBJLDA) approach suggested that the strong spin-orbit coupling (SOC) in zinc blende (ZB) TlN may cause an opening of a narrow gap of 0.11 eV at the $\Gamma$ point of the Brillouin Zone (BZ) \cite{TlInN}, similarly to the case of HgS \cite{HgS1}. Such a feature of the band structure is exhibited by topological insulators \cite{HgS1,HgS2}.

Despite the fact that the Tl$_x$Ga$_{1-x}$N alloys have been already investigated with LDA/GGA methods \cite{TlGaN} some properties of those systems are still unknown. Since the standard DFT-based calculations are insufficient for a prediction of band structures for narrow band-gap systems, the values of the band-gap as well as the SOC-driven separation between the heavy-hole and the split-off band at the $\Gamma$ point of the BZ ($\Delta_{SO}$) require studies with more advanced methods.

In this work, the band structures of ZB Tl$_x$Ga$_{1-x}$N alloys are investigated with the MBJLDA potential \cite{MBJLDA}. This well-known method has already been examined for semiconductors exhibiting narrow band-gaps, e.g. InAs and InSb \cite{PRB_82_205212}, and also used in the former study of Tl$_x$In$_{1-x}$N alloys \cite{TlInN}. The main aim of this work is a prediction of the band-gaps of Tl-doped GaN materials.

\section{Computational details}

The electronic structure calculations for ZB Tl$_x$Ga$_{1-x}$N systems have been performed with the use of the ABINIT code \cite{Abinit1, Abinit2}. First, the equilibrium geometries were found via stresses/forces relaxation for PAW atomic data sets generated with the Atompaw package \cite{Atompaw} with the Perdew-Wang \cite{LDA} parametrization for the exchange-correlation energy. Then, the calculations of the band structures with the use of the norm-conserving fully relativistic pseudopotentials were performed with the MBJLDA (TB09) functional \cite{MBJLDA}. The pseudopotentials were generated  using the Atomic Pseudopotential Engine \cite{APE}. Such a complex calculation scheme was necessary to obtain reasonable results of LDA-predicted structural parameters and MBJLDA band-gaps with SOC effects included.
A supercell consisting of 16-atoms ($2\times2\times2$ multiplicity of primitive fcc cell) was used to simulate the alloy. The most possibly uniform arrangements of atoms were chosen. The total energy convergence in the plane wave basis was sufficient for 50 Ha energy cut-off and the $4\times4\times4$  {\bf k}-point mesh with standard shifts for fcc lattice.

\section{Results and discussion}

The computed lattice parameter {\it a}, band-gap (E$_g$), and split-off energy ($\Delta_{SO}$) for parent compounds, zinc blende GaN and TlN, are gathered in Table \ref{table1}. As seen, the structural relaxation with the LDA potential for GaN leads to very good agreement between the calculated result and the accesible experimental data. In the case of TlN, only theoretical predictions are available in the literature. The calculated here lattice parameter {\it a} of this system is close to other LDA results, being smaller than the GGA one.

It is a well-known fact, that band-gaps of insulators are underestimated by the LDA approach. The obtained (LDA) value of the band-gap for GaN, E$_g$~=~1.69~eV, is almost two times lower than the experimetal one (3.30 eV \cite{GaN_Eg}). The MBJLDA-calculated E$_g$ of 3.04 eV is also slightly too low. It is worth noting that this value is very close to the result of the full-potential calculations, E$_g$~=~3.03~eV, reported in Ref. \cite{InGaN_alchemia}. Therefore, the accuracy of the pseudopotential scheme used here seems to be satisfactory. Furthermore, the obtained in both approaches (LDA/MBJLDA) value of $\Delta_{SO}$ =~16~meV for GaN is in very good agreement with the experimental $\Delta_{SO}$ =~17~meV \cite{GaN_Eg}.

\begin{figure}
\includegraphics[width=8cm]{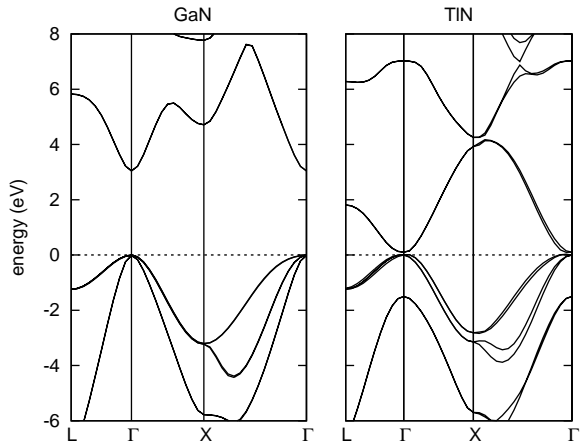}
\caption{Band structures of zinc blende GaN (left panel) and TlN (right panel) calculated within MBJLDA.}
\label{Fig1}
\end{figure}

The MBJLDA band structures for GaN and TlN are presented in Fig.~\ref{Fig1}. One can notice that the overall shape of valence bands is similar in both systems. However, the Tl-based compound exhibits a semimetallic character with an inverted band-gap \cite{TlGaN}, beeing related to the anomalous order of N $2p$ bands, since the quartet N 2$p_{3/2}$ is below the doublet N 2$p_{1/2}$ \cite{TlInN}. The strong SOC in TlN leads to an opening of a narrow gap at the $\Gamma$ point of BZ, estimated to be equal 0.03 and 0.10~eV from LDA and MBJLDA calculations, respectively. The high atomic number of Tl is also reflected in significant splittings of heavy-hole bands, e.g. in the $X-\Gamma$ line of BZ, as can be seen in Fig. \ref{Fig1}.

Because high values of $\Delta_{SO}$ for TlN, ranging from 1.4 to 2.0~eV, have been reported in the literature \cite{TlInN,TlN1,TlN4}, one may consider that the position of the light-hole bands of such a narrow band-gap system is affected by the overestimation of the negative band-gap in the standard LDA/GGA calculations. Thus, the lower MBJLDA value of $\Delta_{SO}$~$\approx$~1.5~eV, seems to be more adequate. It is worth noting that the obtained $\Delta_{SO}$ is slightly higher than the results of former MBJLDA study for the same volume of the unit cell \cite{TlInN}. This discrepancy is related to the fact that in the present study the calculations were performed fully {\it ab initio}. Namely, the parameter $c$ in the MBJLDA exchange-correlation functional \cite{MBJLDA} was not tuned empirically.

\begin{figure}
\includegraphics[width=8cm]{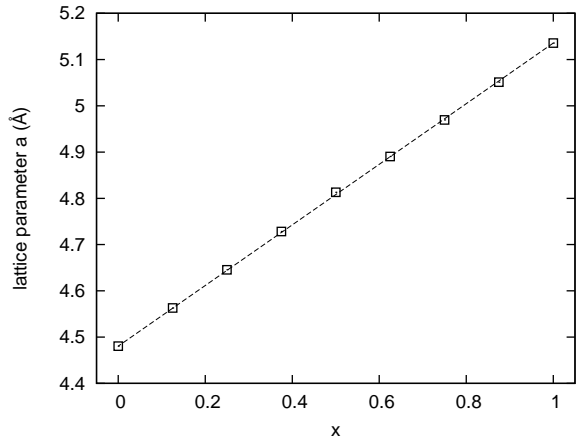}
\caption{Calculated (LDA) lattice parameters {\it a} of zinc blende Tl$_x$Ga$_{1-x}$N alloys as a function of thallium content $x$.}
\label{Fig2}
\end{figure}

The calculated lattice parameters {\it a} for ZB Tl$_x$Ga$_{1-x}$N alloys, presented in Fig. \ref{Fig2}, exhibit a linear behavior (the Vegard's law), similarly to the former results for Tl$_x$Al$_{1-x}$N \cite{TlAlN1} and Tl$_x$In$_{1-x}$N \cite{TlInN} materials. The strong bowing of lattice parameters {\it a} as a function of a Tl content $x$ reported in the former study for Tl$_x$Ga$_{1-x}$N \cite{TlGaN} systems may be connected with the lack of a relaxation of atomic positions during the volume optimisation of the unit cell. Such calculations lead to similar results to those of the virtual crystal approximation, which have been discussed for some nitride systems, e.g. In$_x$Ga$_{1-x}$N \cite{InGaN_alchemia} and AlN$_{1-x}$P$_x$ \cite{AlPN}.

\begin{figure}
\includegraphics[width=8cm]{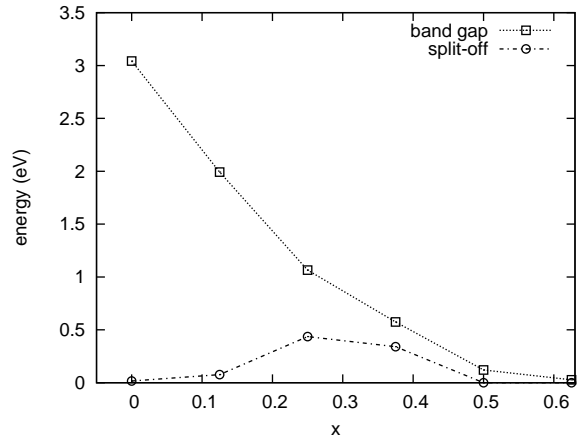}
\caption{Calculated by MBJLDA band-gaps (E$_g$) and split-off energies ($\Delta_{SO}$) for Tl$_x$Ga$_{1-x}$N alloys as a function of thallium content $x$.}
\label{Fig3}
\end{figure}

The dependence of calculated within the MBJLDA approach band-gaps as a function of a Tl content in Tl$_x$Ga$_{1-x}$N alloys, depicted in Fig. \ref{Fig3}, exhibits an almost linear behavior for $x < 0.25$ (for E$_g$ in the range from about 3 to 1~eV). An analogous linear dependence of E$_g$ in Tl$_x$In$_{1-x}$N systems was reported {\it ab initio} \cite{TlInN}, although some group-III nitride semiconductors exhibit a strong band-gap bowing, e.g. In$_x$Ga$_{1-x}$N and In$_x$Al$_{1-x}$N alloys \cite{Gorczyca}.  Interestingly, the experimental studies for Sc-doped GaN and AlN materials also reported a linear behaviour of E$_g$ as a function of a Sc content \cite{ScGaN,ScAlN}. One can consider that the possible solubility of thallium in GaN is rather questionable for high contents of this element in such a material.
Therefore, the linear reduction of E$_g$~ $\approx$ 0.08 eV per Tl  atomic \% was predicted arbitrarily, basing only on data for $x < 0.25$. Since the band-gap of Tl$_{0.5}$Ga$_{0.5}$N is almost zero, the Tl$_x$Ga$_{1-x}$N  systems are predicted to be semimetallic for $x > 0.5$. It is worth noting that the presented here dependence of E$_g$ for Tl$_x$Ga$_{1-x}$N is significantly different from that reported earlier \cite{TlGaN}. Namely, the drop of E$_g$ is less sharp in the LDA results, which cannot be corrected by a simple shift of bands (the so-called 'scissor operator'). Therefore, the band structure of semiconductors doped with heavy elements should be investigated with methods beyond the standard (LDA/GGA) DFT approach, e.g. hybrid exchange-correlation functionals or the used here MBJLDA potential. 

A strong enhancement of SOC-driven splitting between the heavy-hole and split-off band at the $\Gamma$ point of BZ for Tl$_x$Ga$_{1-x}$N alloys, related to the high atomic number of thallium, is expected to be pronounced. However, the results of fully relativistic MBJLDA calculations indicate a significantly weaker influence of Tl atoms on the split-off band in the GaN host systems, i.e. $\Delta_{SO} < 0.5$~eV for $x=0.25$, when compared to the case of Tl$_x$In$_{1-x}$N \cite{TlInN}. Furthermore, for $x > 0.25$, the split-off band is even shifted to the vicinity of the valence band maximum. This effect may be explained by the fact that the semimetallic systems with a higher Tl content exhibit an inverted band-gap. Therefore, the arrangement of particular bands in materials being close to the inversion of the band-gap may be complex and investigations of such an issue are beyond the scope of this work.

\section{Conclusions}

The structural and electronic properties of ZB Tl$_x$Ga$_{1-x}$N alloys were investigated within LDA/MBJLDA calculations. The linear dependence of the lattice parameter {\it a} as a function of a Tl content was revealed. The singnificant decrease of the band-gap of Tl-doped GaN materials is also predicted as a linear function of a Tl content $x$ for $x<0.25$, whereas the semimetallic character is expected for $x>0.5$. The obtained SOC-driven splittings between the heavy-hole and split-off band at the $\Gamma$ point of BZ in Tl$_x$Ga$_{1-x}$N alloys are relatively low when compared to those of Tl-doped InN systems.

\section*{Acknowledgments}
The calculations were performed in Wroc\l aw Center for Networking and Supercomputing (Grant No. 158).

\begin{table}
\caption{Lattice parameter {\it a}, band-gap (E$_g$), and split-off energy ($\Delta_{SO}$) of zinc blende GaN and TlN.}
\label{table1}
\begin{tabular}{llll}
 &  {\it a} (\AA) &  E$_g$ (eV) & $\Delta_{SO}$ (eV) \\ \hline
GaN: & & &  \\
this work LDA & 4.481 & 1.69 &  0.016 \\
this work MBJLDA & - & 3.04 &  0.016 \\
LDA Ref. \cite{InGaN_alchemia} & 4.49 & - & - \\
MBJLDA Ref. \cite{InGaN_alchemia} & - & 3.03 \\
LDA Ref. \cite{SSC_116_421} & - & - & 0.019 \\
exp. Ref. \cite{GaN_a} & 4.49 & - & - \\
exp. Ref. \cite{GaN_Eg} & - & 3.30 & 0.017 \\
TlN: & & &  \\
this work LDA & 5.136 & 0.03 & 2.078 \\
this work MBJLDA & - & 0.10 & 1.499 \\
MBJLDA Ref. \cite{TlInN} & - & 0.11 & 1.397 \\
LDA Ref. \cite{TlN1} & 5.133 & - & 2.000 \\
GGA Ref. \cite{TlN4} &  5.153 & - & 1.830 \\
\end{tabular}
\end{table}

\end{document}